\documentclass[%
reprint,
 amsmath, amssymb,
 aps,
]{revtex4-2}

\usepackage{graphicx}
\usepackage{dcolumn}
\usepackage{bm}
\usepackage{algorithm}
\usepackage{algpseudocode}
\usepackage{placeins}
\usepackage{soul}
\usepackage{appendix}
\usepackage{chngcntr}
\usepackage{todonotes}

\let\OLDthebibliography\thebibliography
\renewcommand\thebibliography[1]{
  \OLDthebibliography{#1}
  \setlength{\parskip}{2pt}
  \setlength{\itemsep}{2pt plus 0.3ex}
}

\begin{document}

\preprint{}

\title{Simple synthetic molecular dynamics for efficient trajectory generation}
\author{John D. Russo}
\author{Daniel M. Zuckerman}
 \email{zuckermd@ohsu.edu}
\affiliation{%
 Department of Biomedical Engineering,
Oregon Health and Science University, Portland, OR
}

\date{\today}

\begin{abstract}
Synthetic molecular dynamics (synMD) trajectories from learned generative models have been proposed as a useful addition to the biomolecular simulation toolbox.
The computational expense of explicitly integrating the equations of motion in molecular dynamics currently is a severe limit on the number and length of trajectories which can be generated for complex systems.
Approximate, but more computationally efficient, generative models can be used in place of explicit integration of the equations of motion, and can produce meaningful trajectories at greatly reduced computational cost.
Here, we demonstrate a very simple synMD approach using a fine-grained Markov state model (MSM) with states mapped to specific atomistic configurations, which provides an exactly solvable reference.
We anticipate this simple approach will enable rapid, effective testing of enhanced sampling algorithms in highly non-trivial models for both equilibrium and non-equilibrium problems.
We demonstrate the use of a MSM to generate atomistic synMD trajectories for the fast-folding miniprotein Trp-cage, at a rate of over 200 milliseconds per day on a standard workstation.
We employ a non-standard clustering for MSM generation that appears to better preserve kinetic properties at shorter lag times than a conventional MSM.
We also show a parallelizable workflow that backmaps discrete synMD trajectories to full-coordinate representations at dynamic resolution for efficient analysis.
\end{abstract}

\maketitle

\section{Introduction}

\begin{figure}
    \centering
    \includegraphics[width=0.95\linewidth]{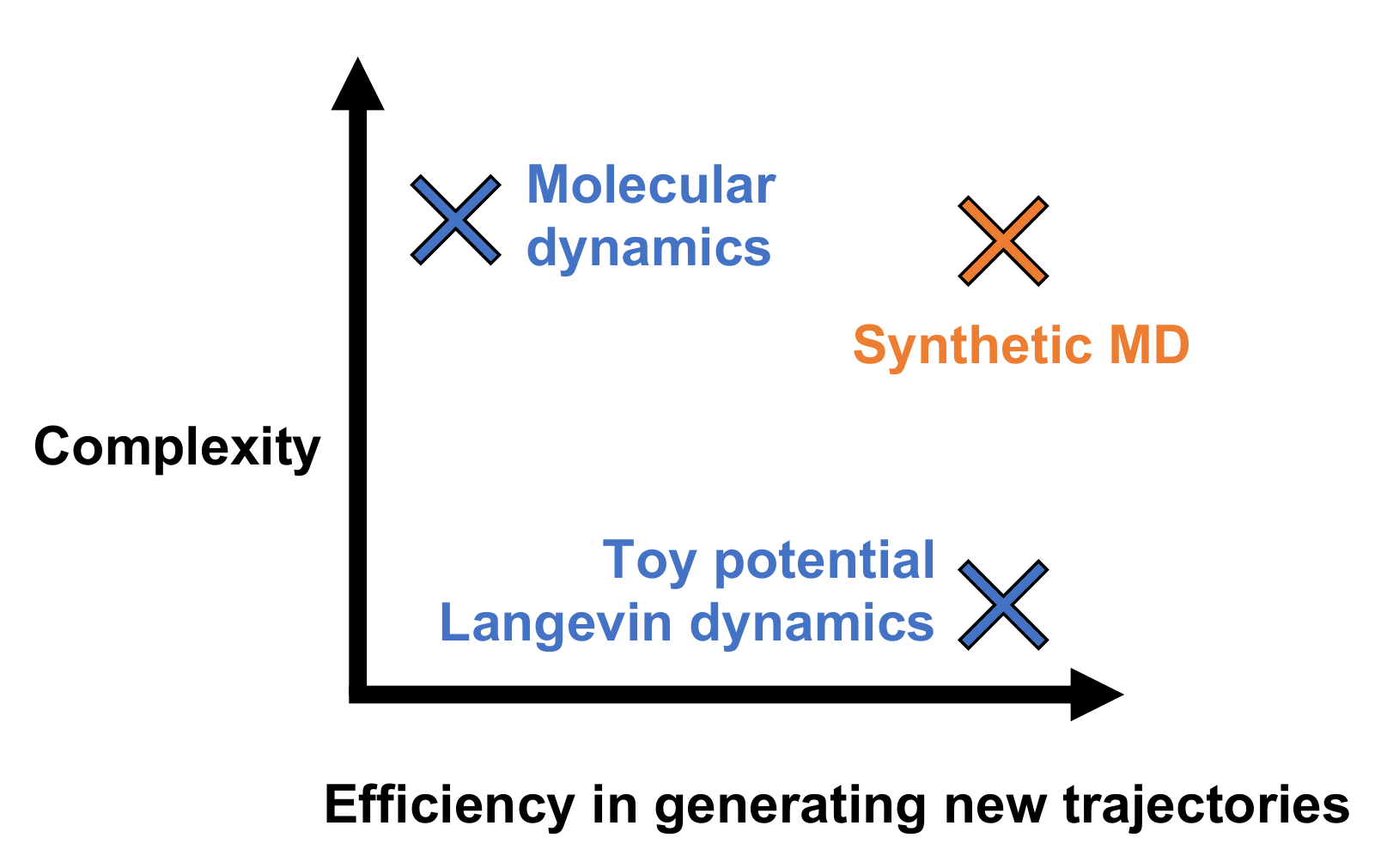}
    \caption{Comparison of different simulation methods.
    Molecular dynamics can simulate highly complex systems, at the cost of great computational expense.
    Simpler toy potentials simulated under, for example, Langevin dynamics, can be highly efficient but lack the complexity present in real systems.
    }
    \label{fig:efficiency}
\end{figure}

The overall goals of molecular dynamics (MD) simulation are to generate sufficiently well-sampled and accurate trajectories, but these are hindered by notable challenges.
On the one hand, inadequate sampling of complex systems prevents complete characterization of force-field accuracy, impeding the improvement of force field models.
On the other, poor sampling also complicates the development of new sampling methods, because it is effectively impossible to gauge the success of a new method without reference \emph{simulation} data.
Well-sampled simulation data (rather than experimental data) on complex systems is required as a reference for methods development because even perfectly sampled models are not expected to agree with experiments, again because of model inaccuracy \cite{shaw_fastfolding}.

Synthetic molecular dynamics (synMD), i.e., the generation of approximate but arbitrarily long trajectories of highly complex models \cite{cg-gml, rnn-aimd, deepgen-msm, learning-effective-dynamics, ferguson-latent, tiwary-lstm, tiwary-rnn-sampling}, can directly aid methods development for sampling and hence indirectly contribute to force field development.
In the long term, increasingly accurate synMD models may ultimately provide a partial replacement for standard MD.

The limitations of conventional MD simulation for biomolecules are well known.
Record millisecond-timescale simulations are only achievable for relatively small and simple proteins, even with substantial computational resources \cite{shaw_fastfolding,shaw_benchmarks}.
In contrast, more complex processes of biological interest in larger systems span timescales up to to seconds and beyond \cite{covid1,covid2,biological-timescales}, which are inaccessible by conventional MD \cite{zwier2010, md-review}. 

MD limitations have motivated the development of numerous alternative strategies.
Coarse-graining atoms using a force field such as MARTINI \cite{martini}, or representing the solvent with an implicit  model \cite{implicit-book} are strategies for accelerating simulation speed by reducing the number of atoms being simulated, as are statistical mechanics-based coarse-graining strategies like force-matching \cite{force-matching,md-cg-ml,md-ml}.
Enhanced equilibrium sampling methods such as replica exchange, metadynamics, or umbrella sampling with weighted histogram analysis employ modified energy landscapes, and are popular alternatives to conventional MD for atomistic systems \cite{metadynamics,replica-exchange,sampling-review,wham,umbrella}.
Path-sampling methods, including weighted ensemble and forward-flux sampling among others, 
aim to improve simulation efficiency by focusing computational resources on regions of interest and can provide unbiased non-equilibrium observables.
\cite{milestoning,tessellation-milestoning,westpa,forwardflux,noneq-umbrella,noneq-umbrella-dickson,noneq-ss-sampling,transition-interface-sampling,transition-interface, huber}. 
Finally, Markov state models (MSMs) are a useful tool for connecting data from independent simulations which sample different but overlapping regions of phase space \cite{msm-noe, msm-noe-2}.

Despite this significant progress, a persistent challenge in methods development for biomolecular simulation  -- which remains ongoing for essentially all of the strategies noted above -- is the lack of validation data, i.e., extremely well-sampled MD data for systems of interest.
As illustrated in Fig.~\ref{fig:efficiency}, well-sampled MD runs are typically slow for complex systems.
This makes them infeasible for use as a step in methods development pipelines because sufficiently complex systems generally cannot be sampled well enough to provide reference values for comparison.
Simpler and faster systems such as low-dimensional potentials likely will not capture sufficient complexity to challenge the methods being tested.

Here, we describe a simple synthetic MD workflow based on MSMs, in which a generative model is trained using a set of initial, standard molecular dynamics data. 
MSMs \cite{msm-noe,msm-noe-2,msm-amaro, msm-short-offequilibrium}, with states mapped to specific atomistic configurations, are perhaps the simplest type of generative model.
MSM variants such as history-augmented Markov state models (haMSMs) and other MSM alternatives can also be used \cite{msm-computational, msm-beyond, msm-unbiased, msm-dga, msm-memory, kemeny, markov-renewal, msm-observable-operator, tessellation-milestoning} again by mapping discrete states to specific configurations.
A special class of coordinate-generative MSMs can also be used to probabilistically generate new, out-of-sample structures \cite{deepgen-msm}.
Our work is distinguished from notable previous synMD efforts \cite{deepgen-msm, learning-effective-dynamics,tiwary-lstm,tiwary-rnn-sampling,ferguson-latent,rnn-aimd,cg-gml} by its simplicity and the availability of exact solutions.

In this preliminary work, we build a detailed generative MSM from folding trajectories of the Trp-cage miniprotein \cite{shaw_fastfolding}.
We employ a simple stratification strategy to augment the usual MSM clustering that appears to preserve kinetic characteristics at smaller lag times than might otherwise be necessary for validation \cite{suarez-msm}.
We generate synMD trajectories at a rate of $\sim$250 ms/day on a MacBook computer, compared to $\sim$100 $\mu$s/day on the Anton supercomputer for the original trajectories.
We confirm that the synMD trajectories reproduce observables of the MD training data consistent with known capabilities and limitations of MSMs \cite{suarez-msm}, and that the synMD trajectories replicate exactly calculable equilibrium and kinetic properties of the MSM as expected.
We also demonstrate dynamic resolution analysis of the synMD trajectories, where full-coordinate structures are only backmapped within time intervals and at a time-resolution of interest, rather than to each generated point, enabling more efficient analysis.


\section{Methods}

\begin{figure}
    \centering
    \includegraphics[width=0.5\linewidth]{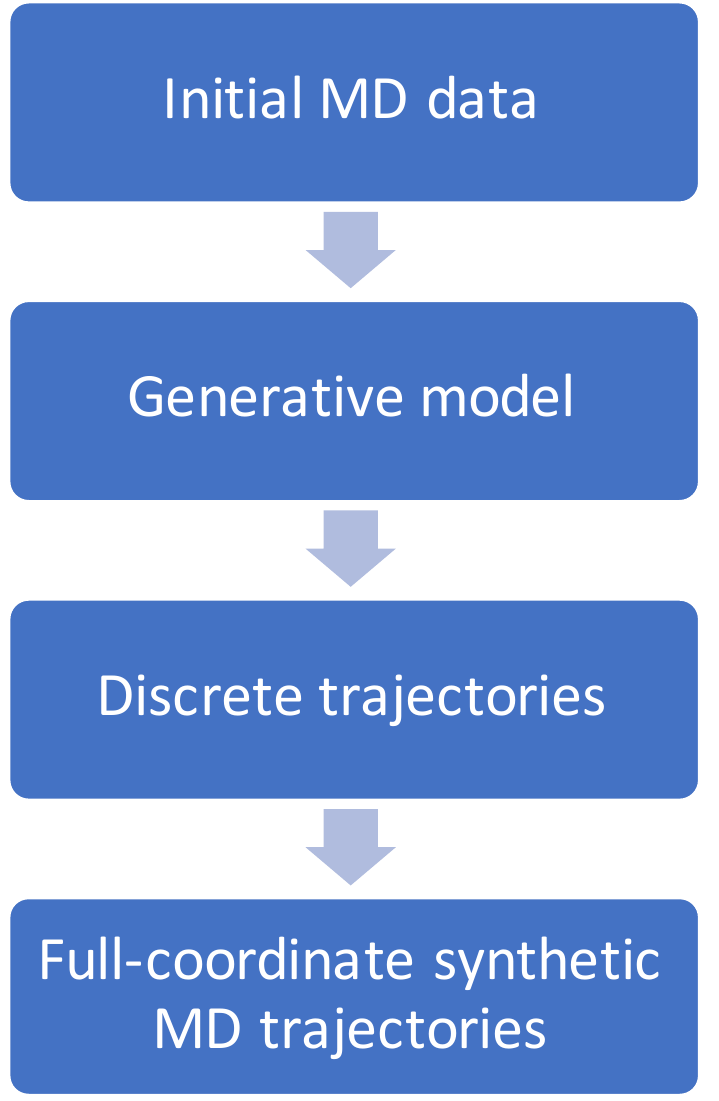}
    \caption{The synthetic MD workflow using discrete-state models. 
    Initial MD simulation data is used to construct a discrete generative model.
    Discrete state trajectories are efficiently generated from this model, and back-mapped to full-coordinate structures.
    This last step is trivially parallelizable.}
    \label{fig:flow}
\end{figure}

\begin{figure*}
    \centering
    \includegraphics[width=\textwidth]{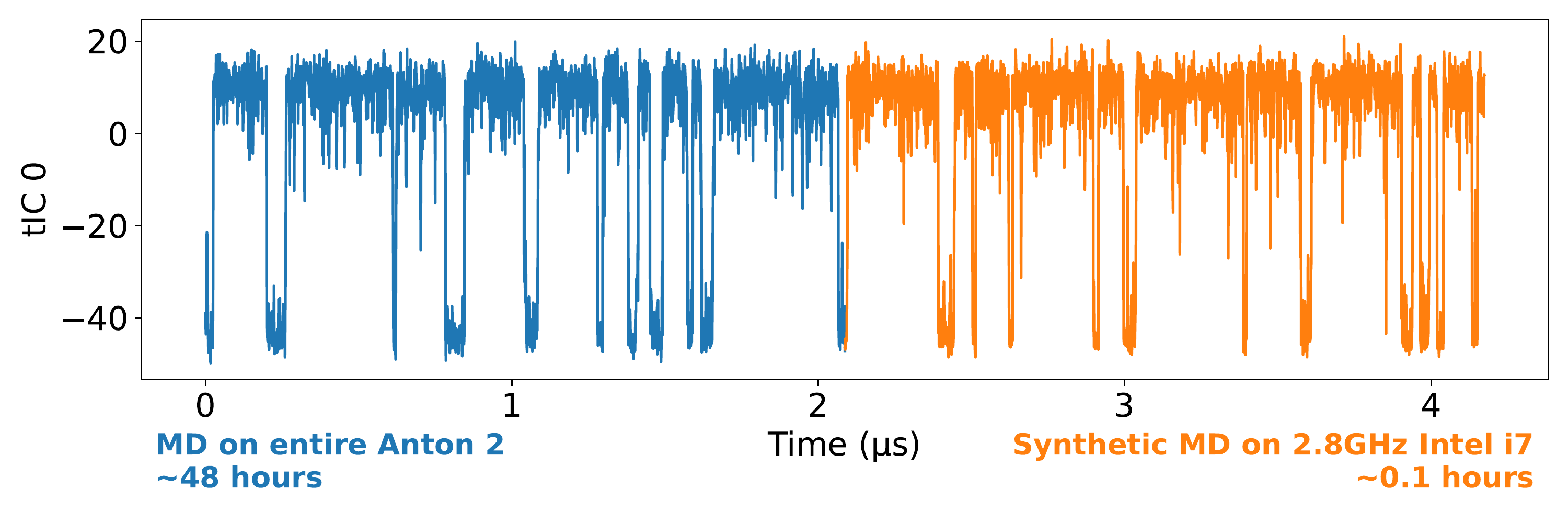}
    \caption{
    Original 208 µs trajectory from MD simulations of the protein Trp-cage \cite{shaw_fastfolding}, extended with another 208 µs of synthetic MD. 
    The synthetic MD trajectory was constructed according to Sec.~\ref{sec:workflow} at 10 ns resolution, initialized from the final point of the MD trajectory.
    The synthetic trajectory is projected here into the same tICA space computed from the MD trajectory for consistency.
    Only the first tIC, which strongly contrasts the folded and unfolded states, is shown.
    }
    \label{fig:trajectory}
\end{figure*}

\subsection{Workflows}\label{sec:workflow}
We present two main workflows for producing synthetic MD trajectories.
First, we describe a generic strategy for efficiently generating trajectories with full atomic coordinates.
Second, we outline a strategy to efficiently generate extremely long atomistic trajectories at a coarse temporal resolution, followed by enhancement of the resolution in post-processing for time intervals of interest.

In the standard synthetic MD workflow employing discrete states (Fig.~\ref{fig:flow}), a generative model employing a discretization of configuration space, such as an MSM, is first built from an initial set of traditional MD trajectories \cite{shaw_fastfolding}.
A specific full-coordinate atomistic configuration is associated with each discrete state.
The generative model is then used to simulate trajectories, which will be time-ordered lists of discrete configurational states, stored as integers.
Discrete trajectory generation typically will be an extremely rapid process.
These trajectories are then back-mapped to the saved atomic coordinate structures.
Because the discrete trajectories are generated before assigning full-coordinate structures, the back-mapping is highly parallelizable.
Finally, the full-coordinate trajectory is written to disk in a standard MD format, enabling processing by standard tools.

\begin{figure}  
    \centering
    \includegraphics[width=1.1\linewidth]{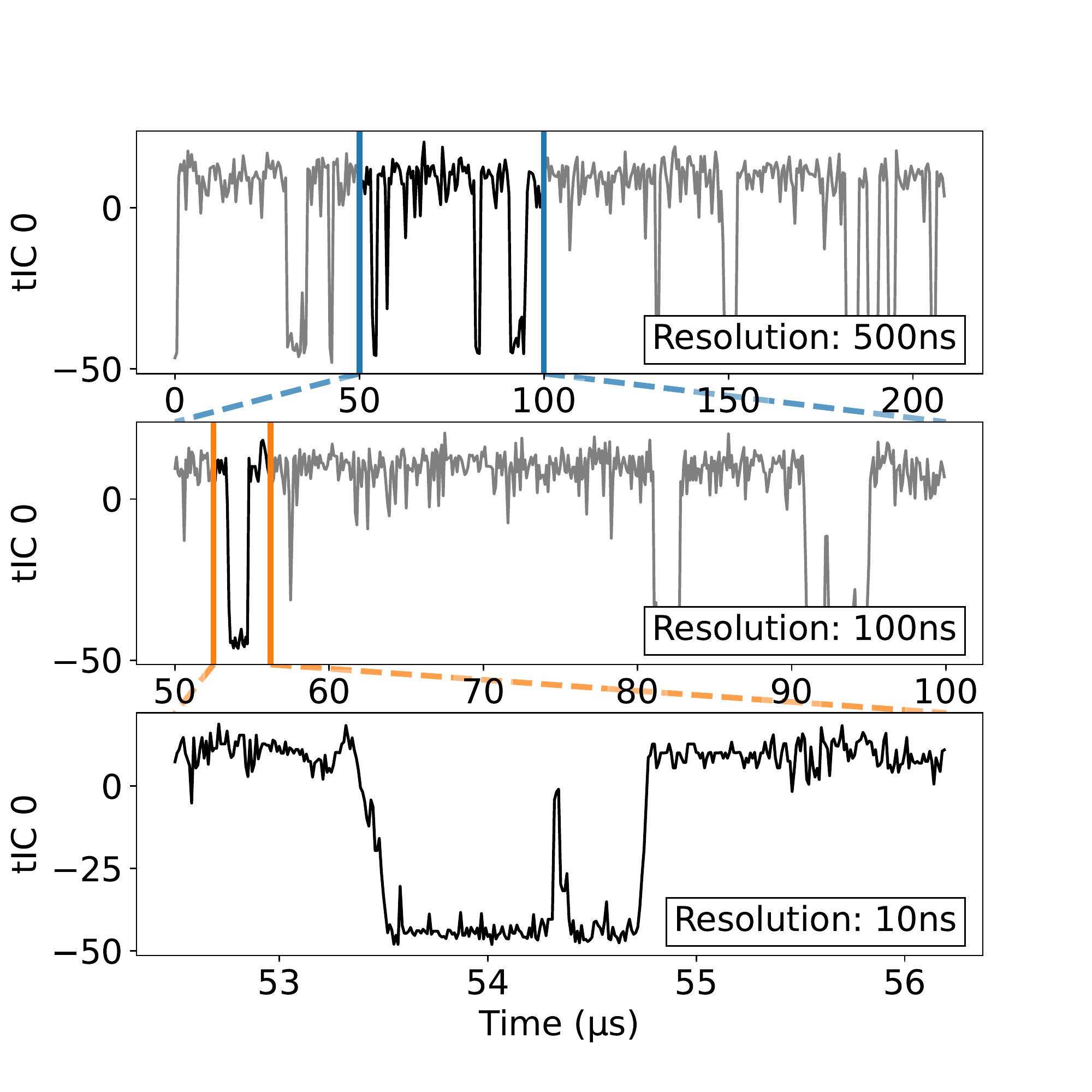}
    \caption{
    Synthetic MD trajectory for the protein Trp-cage, shown at varying levels of temporal resolution obtained in post-analysis. 
    The full-coordinate trajectory may be initially back-mapped over only subsampled points from the generated discrete trajectory (top). 
    Intervals of interest can later be backmapped at higher resolution (middle and bottom).
    The first tIC (time-independent component) is taken as a representative coordinate because it clearly shows folding transitions.  SynMD trajectories include all atomistic coordinates, enabling arbitrary analysis.
    }
    \label{fig:zoom}
\end{figure}

Synthetic MD also enables a dynamic resolution workflow (Fig. \ref{fig:zoom}), where very long trajectories can be efficiently generated, and enhanced temporal resolution added to regions of interest in post-processing.
In this workflow, the generative model is used to build a very long discrete trajectory.
However, only a temporally subsampled set of points from the discrete trajectory are back-mapped to full-coordinate atomistic configurations, rather than the full trajectory.
This enables ``telescoping'' detailed analysis of long trajectories that would be infeasible at full temporal resolution because of the large number of snapshots generated in synMD.

\subsection{Simple generative model: MSM of Trp-cage}\label{sec:model}
To demonstrate the synMD approach, we employed a nearly standard MSM as a generative model, built with pyEMMA \cite{scherer_pyemma_2015}.
The clustering described below is slightly different than for typical MSMs.
The original MD trajectory from a 208 $\mu$s simulation of the protein Trp-cage  \cite{shaw_fastfolding} was first featurized with residue-residue minimum RMSD, excluding nearest neighbors.
Next, tICA dimensionality reduction was performed at a 10ns lag time with 10 tICs,  using commute maps for eigenvector scaling.

The dimensionality-reduced trajectories were clustered using a stratified k-means approach, which differs somewhat from typical MSM workflows.
A coordinate of interest is first stratified into bins, and then k-means clustering is independently performed in each bin.
Stratification guarantees an even distribution of states along coordinates of interest.
In this case, we stratified along tIC 0, which sharply distinguishes the folded and unfolded states, guaranteeing reasonable coverage of transition regions in this coordinate.
With 20 k-means centers for each stratified bin, there were a total of 1020 clusters which form the discrete states of the generative model.

The discretized trajectories were used to build a MSM at a 10ns lag time, chosen to balance time resolution with reasonable kinetic fidelity \cite{suarez-msm}. 
The MSM was symmetrized to ensure satisfaction of detailed balance by adding the count matrix to its transpose.
For each discrete state, a single representative structure was randomly chosen from all structures assigned to that state.

Some of the choices made in constructing this MSM may decrease model fidelity to the MD training data, but we emphasize our initial goal is to construct a generative model with protein-like complexity to enable downstream analysis and testing.
Indeed, MSMs have fundamental limitations that have been discussed in detail \cite{suarez-msm}.

For reference, we note this MSM produced mean first-passage times (MFPTs) of 12.7$\mu$s for folding and 2.8$\mu$s for unfolding as calculated from the transition matrix using pyEMMA \cite{scherer_pyemma_2015}.

\section{Results}

Five 208 µs trajectories were produced at 10 ns resolution by propagating randomly chosen initial states using the Trp-cage generative model.
This took 5 minutes 41 seconds in total for all five trajectories using a MacBook Pro with a single 2.8GHz Intel i7 processor.
One such trajectory is shown in Fig.~\ref{fig:trajectory}, along with the original MD trajectory.

\begin{figure}
    \centering
    \includegraphics[width=1.0\linewidth]{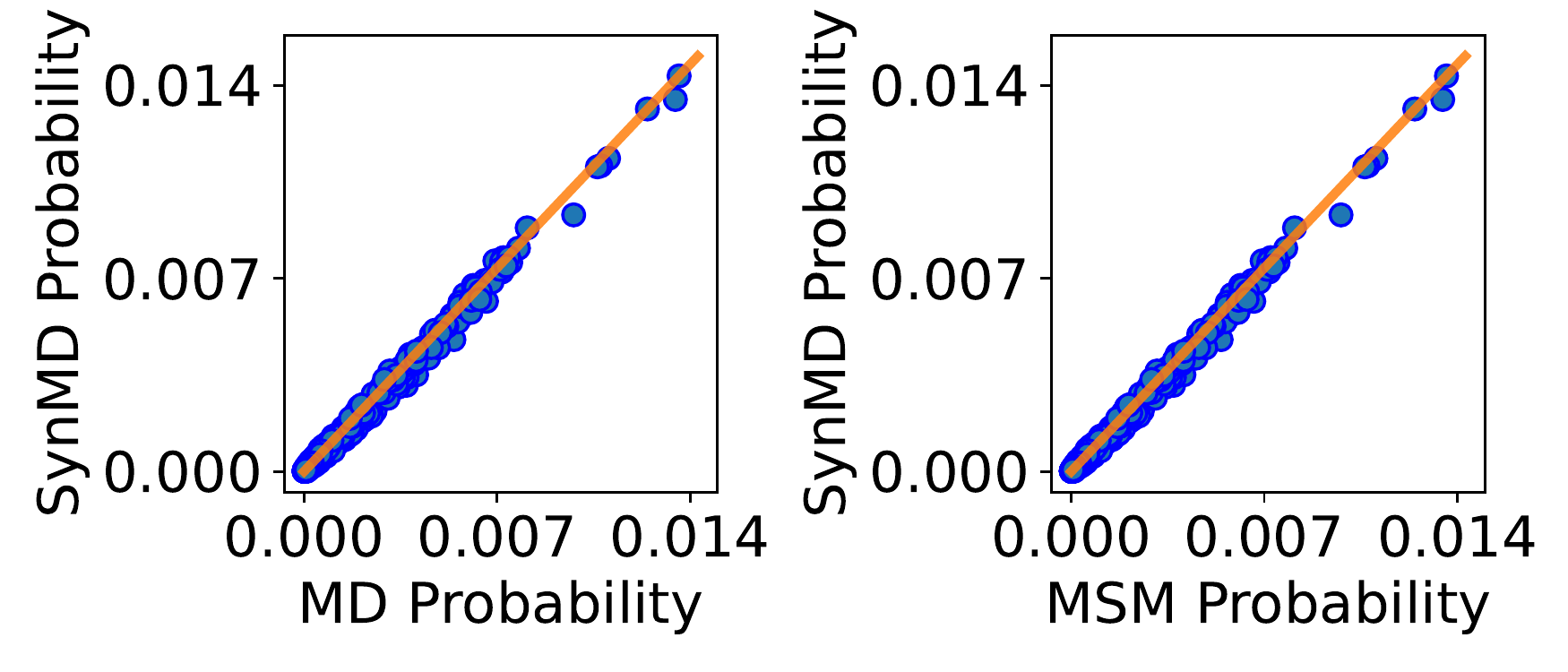}
    \caption{
    Comparison of synMD equilibrium distributions to the MD training data and the generating MSM.
    Each point represents the fractional occupancy of a discrete state of the MSM, with MSM values computed from the stationary distribution of the transition matrix.
    SynMD values are averages over the five 208 $\mu$s synMD trajectories.
    }
    \label{fig:distributions}
\end{figure}

Analysis of these trajectories' equilibrium distributions is consistent with the original MD trajectory data, as well as the underlying MSM, as shown in Fig.~\ref{fig:distributions}.
Likewise, the MFPT values estimated from the synMD trajectories were $4.1 \pm 1.5$ $\mu$s for unfolding and 
$18.3 \pm 9.6$ $\mu$s for folding, consistent with the reference values of 2.8 $\mu$s and 12.7 $\mu$s computed directly from the MSM transition matrix.

\section{Discussion}

We have explored a very simple approach to generating synthetic molecular dynamics (synMD) trajectories based on Markov state models (MSMs), with the motivation of rapidly generating trajectories in highly non-trivial systems that can be solved exactly, in turn providing ideal test beds for methods development.
Previous work has employed a range of deep-learning techniques \cite{deepgen-msm, learning-effective-dynamics,tiwary-lstm,tiwary-rnn-sampling,ferguson-latent,rnn-aimd,cg-gml}.

We show that MSM-based synMD trajectories are generated at multiple orders of magnitude speedup over conventional MD, and confirm that the trajectories reproduce exactly-solvable equilibrium and kinetic properties of the generative model.
Our generative MSM was able to employ a shorter lag time -- providing higher mechanistic resolution \cite{suarez-msm} -- because of an apparently novel stratified approach to state clustering.

Rapid generation of synMD trajectories should be very useful in testing new methods because it provides arbitrary amounts of data in highly complex, but exactly solvable models.
Such a framework could be particularly valuable for path sampling, enabling careful estimation of variance based on different choices of hyper-parameters.
SynMD can also advance methods development for trajectory analysis tools \cite{msm-unbiased,nonparametric} based on controlled amounts of synMD data, mimicking the low-data regime typical for MD trajectory sets.
Even MSM analysis protocols can be tested using synMD based on a fine-grained MSM, so long as the MSM used for analysis is blinded to the fine-grained MSM used to generate trajectories.
SynMD may also be useful for generating an arbitrary number of stochastic mechanistic pathways encoded by the generative model, which may be compared to experimental or higher-quality simulated data to further refine the generative model \cite{sugita2018refining}.

It is feasible to construct significantly improved generative models within the MSM framework.
For example, much finer-grained states can be employed, and established adaptive approaches for selecting key regions for further simulation (of MD training data) are available \cite{clementi-adaptive1,clementi-adaptive2,pande-msm-1,pande-msm-2}.
Training data from polarizable or hybrid quantum/classical force fields could be used to refine a conventional MSM as needed.
Numerous MSM-like discrete-state models have been developed incorporating more dynamical information -- i.e., trajectory history -- than conventional MSMs 
\cite{msm-computational, msm-beyond, msm-unbiased, msm-dga, msm-memory, kemeny, markov-renewal, msm-observable-operator, tessellation-milestoning}.
For example, haMSMs are unbiased for kinetics at any lag time and were shown to significantly outperform conventional MSMs in characterizing mechanistic details of protein folding \cite{suarez-msm,msm-unbiased}.
Deep generative MSMs can be used to stochastically generate new out-of-sample structures \cite{deepgen-msm}.

More modern machine learning strategies will undoubtedly continue to play a large role in synMD.
Frameworks such as variational autoencoders 
and recurrent neural networks including long short-term memory neural networks \cite{tiwary-rnn-sampling,tiwary-lstm,rnn-aimd} have led to models with an improved ability to generate MD-like discrete-state trajectories; note that current MSMs and variants have not been optimized for this task, which is critical to synMD.
Mixture density network autoencoders \cite{learning-effective-dynamics} and latent space simulators \cite{ferguson-latent} generate trajectories in a lower-dimensional continuous space, and provide a mapping to full-coordinate representations.

\begin{acknowledgments}
We appreciate valuable input from Jeremy Copperman, and are grateful for support from the NIH via Grant GM115805 and the NSF via Grant MCB 2119837.  We thank Pratyush Tiwary and Andrew Ferguson for useful guidance on the current literature.

\end{acknowledgments}

\bibliography{bibliography}

\end{document}